\documentstyle[12pt,aasms]{article}
\tightenlines

\newcommand{\bq}{\begin{equation}}
\newcommand{\eq}{\end{equation}}

\begin{document}
 
\title{Determining the Motion of the Local Group\\Using SN Ia Light Curve Shapes}

\author{Adam G. Riess,  William H. Press, Robert P. Kirshner}
\affil{Harvard-Smithsonian Center for Astrophysics, 60 Garden Street, Cambridge, MA 02138}

\begin{abstract}
We have measured our Galaxy's motion relative to distant galaxies in
which type Ia supernovae (SN Ia) have been observed.  The effective
recession velocity of this sample is 7000 km s$^{-1}$, which
approaches the depth of the survey of brightest cluster galaxies by
Lauer and Postman (1994).  We use the Light Curve Shape (LCS) method
for deriving distances to SN Ia, providing relative distance estimates
to individual supernovae with a precision of $\sim$ 5\% (Riess, Press,
\& Kirshner 1995).  Analyzing the distribution on the sky of velocity
residuals from a pure Hubble flow for 13 recent SN Ia drawn primarily
from the Cal\'{a}n/Tololo survey (Hamuy 1993a, 1994, 1995a, 1995b,
Maza et al. 1994), we find the best solution for the motion of the
Local Group in this frame is 600 $\pm 350$ km s$^{-1}$ in the
direction b=260$\deg$ {\it l}=+54$\deg$ with a 1 $\sigma$ error
ellipse that measures 90$\deg$ by 25$\deg$.  This solution is
consistent with the rest frame of the cosmic microwave background
(CMB) as determined by the Cosmic Background Explorer (COBE)
measurement of the dipole temperature anisotropy (Smoot et al. 1992).
It is inconsistent with the velocity observed by Lauer and Postman.

\end{abstract}
subject headings: 
supernovae:general ; cosmology: distance scale and observations; large-scale
structure of universe; Local Group
\vfill\eject

\section{Introduction} 

  The
motions of galaxies provide a dynamical measure of the way mass and light are
distributed in the Universe. In 1976, Rubin et al (Rubin et al. 1976, Rubin
1977)  used giant Sc spiral galaxies as distance indicators to find evidence
for a 600 km s$^{-1}$ coherent motion of neighboring galaxies out to 3500-6500
km s$^{-1}$.  Many subsequent attempts to measure motion that departs from
smooth expansion combine redshifts with distance measures (Burstein 1990). 
This approach yields detections of local flows attributed to nearby mass
concentrations (Virgo, Hydra-Centaurus, Great Attractor) of substantial scale
and amplitude.  The most disturbing finding is the absence of convergence to
the cosmic microwave background (CMB) frame.  Recently, Lauer and Postman
(1994) (LP) surveyed brightest cluster galaxies from Abell clusters and
inferred that, remarkably,  a volume of space out to 8,000-11,000 km s$^{-1}$
is moving at 560 km s$^{-1}$ with respect to the CMB frame.  This lack of
convergence to the CMB frame is hard to understand in conventional pictures of
structure formation (Strauss et al. 1994) and invites investigation by
independent methods.  This requires a bright distance indicator that is useful
out to 10,000 km s$^{-1}$ and which is precise enough to detect velocity
residuals of order 600 km s$^{-1}$.

  The narrow luminosity distribution of
Type Ia supernovae (SN Ia) suggests they can provide deep and accurate distance
estimates, and information in the light curve shape has improved the usefulness
of these objects for cosmological measurements (Phillips 1993, Hamuy et al
1995a).  We have developed statistical tools to derive distances using distance
independent information contained in supernova light curves (Riess, Press,
Kirshner 1995 hereinafter RPK).  Following Phillips (1993), the method relies
on a ``training set'' of light curves for supernovae in galaxies with accurate
relative distance measurements.  This determines the correlation between
luminosity and light curve shape.  When applied to an independent set of 13
distant SN Ia light curves drawn primarily from the Cal\'{a}n/Tololo survey
(Hamuy 1993a, 1994, 1995a, 1995b, Maza et al. 1994, Ford et al. 1993, Riess et
al. 1995), the Light Curve Shape (LCS) method provides more precise distance
estimates than the standard candle assumption, reducing the dispersion around a
Hubble line from 0.5 mag to 0.18 mag.  The LCS method also predicts individual
distance uncertainties: the median distance error for our 13 supernovae is 4\%,
with the rest of the dispersion accounted for by random and bulk flow
velocities.  Although our sample is small, the distances of the galaxies
hosting the supernovae are relatively large (up to cz=30,000 km s$^{-1}$), and
the LCS distances are precise enough to provide information about the velocity
residuals from a pure Hubble flow.    

\newpage
  
\section{Measuring the Local Group
Motion with SN Ia's}    

    We fit a Hubble line to 13 supernovae (analyzed
using LCS, see table 1) to determine simultaneously $H_0$ (which we discuss
elesewhere, see RPK) and three Cartesian velocity components of the mean rest
frame of the host galaxies with respect to the Local Group.  Transformation
from heliocentric redshifts to the Local Group rest frame is done by addition
of the vector ($-30$, $297$, $-27$) km s$^{-1}$ in Galactic Cartesian
coordinates (de Vaucouleurs et al 1991,Lynden-Bell \& Lahav 1988).

For comparing dipole hypotheses, we use a $\chi^2$ statistic which
takes into account both distance and velocity departures from the Hubble line. 
Our individual distance uncertainties come from the LCS method where the errors
inferred are found to be statistically sound (RPK).  Individual supernovae have
motions that are not modeled by a pure Hubble flow plus a motion of the Local
Group.  That galaxies have random velocities (RV's) has long been recognized in
the literature and is evident in the velocity residuals of our sample. 
Adopting RV anywhere in the range of 275 to 800 km sec$^{-1}$ yields the
expected range of $\chi^2$ for $9$ degrees of freedom (13 less 3 velocity
components and H$_o$), with little sensitivity to the value of RV.  We adopt
RV=470 km s$^{-1}$ which yields a $\chi^2$ per degree of freedom of 1 and is
consistent with 400 $\pm139 $ km s$^{-1}$, derived from the recent pairwise
velocity difference estimate from the CFA redshift survey (Marzke et al. 1995).

     To assess how our conclusions depend on our distance and velocity
error model, we investigated other plausible error models.   These consisted of
adopting Marzke's value RV=400 km s$^{-1}$ and then adding in quadrature a term
proportional to the velocity (distance), $\sigma_*$, to represent uncertainty
in photometry, absorption, K corrections, or cosmic scatter; $ \sigma_{\mu}^2 =
\sigma_{LCS}^2 + \sigma_*^2$.  The largest amount of uncertainty one can add
and still maintain a significant $\chi^2$ is 7\% of the velocity
($\sigma_*$=.15 mag), and even this has little effect on the conclusions we
draw.

  Minimizing $\chi^2$ with respect to the four free parameters
determines the best values: $H_0=66$  km s$^{-1}$ Mpc$^{-1}$, $V_x=-90$ km
s$^{-1}$ , $V_y=-510$ km s$^{-1}$ and $V_z=710$ km s$^{-1}$ .  Our calibration
of the SN Ia peak magnitude, derived from the HST Cepheid distance to SN 1972E
(RPK, Sandage et al. 1994), affects only our estimate of $H_0$ and not the
velocity components.  The weighted mean recession velocity of our sample is
7,000 km s$^{-1}$.  To visualize the dipole present in our sample, Figure 1
plots the residual recession velocity from the best Hubble fit (in the velocity
rest frame of the Local Group) versus the location on the sky for each
supernova.  The systematic pattern of residuals which is evident in this figure
clearly shows our motion with respect to these galaxies.

Joint confidence
regions for the four fitted parameters at various
confidence levels can be
determined, in the standard way (Avni 1976, Press et al. 1992), as
ellipsoids
whose boundaries correspond to specified values for
$\Delta\chi^2$ .  In this
case, we determine joint confidence regions for the three velocity
parameters
$V_x,V_y,V_z$, with the fourth parameter ($H_0$) chosen to
minimize
$\chi^2$ for each value of $(V_x,V_y,V_z)$.  With three degrees
of
freedom in the resulting joint confidence region, the boundary of the
68\%
confidence region lies at $\Delta\chi^2=3.5$, the $95\%$ confidence region lies
at $\Delta\chi^2=8.0$, and the boundary
of the 99\% confidence region lies at
$\Delta\chi^2=11.3$. 

We can now ask where the COBE rest frame, and the LP
rest
frame, lie with respect to these confidence regions:   

When the COBE
frame ($10,-542,300$)  km s$^{-1}$ is used $\chi^2=12.5$ (D.F.=9), so
$\Delta\chi^2=3.5$.  This is not only inside the 95\% confidence region, it is
inside the 68\% confidence region, showing that our measurement of the Local
Group motion is highly consistent with the COBE dipole. For the LP frame
($-470,-399,-333$) km s$^{-1}$, $\chi^2=32.7$ (D.F.=9), so $\Delta\chi^2=23.7$.
 This lies outside the 99\% confidence region.  In fact, it is ruled out at 
greater than the $99.99\%$ (``greater than 3-sigma'') confidence level, i.e.
there is less than a 1 in 10,000 chance that our result is an unusual data
sample from the LP rest frame.  We can also determine our consistency with the
LP measurement as well as the LP rest frame by adding our errors in quadrature
with theirs and evaluating the confidence level with which we can say the
difference between the two measurements is zero.  Including the errors of both
measurements yields a reduced rejection level of $99.45\%$.  Alternatively, we
find that the most likely dipole vector for our measurement and the LP
measurement taken jointly is ($-340,-420,140$) with a joint likelihood of 0.7\%
or a mutual rejection of 99.3\%.  Additionally, we have compared our
measurement with the IRAS dipole, deduced by peculiar accelerations, and found
a high level of consistency ($\Delta\chi^2=1.2$, Rowan-Robinson et al 1990). 
Finally, we can reject the null hypothesis, i.e. no net motion with respect to
the supernova frame, at the 99\% confidence level. 

Our ability to exclude the
LP  
rest frame at interesting confidence levels derives from the
small
dispersion of the LCS method, yet the same trend is present at lower
precision if the
analysis is redone assuming SN Ia's to be identical standard
candles with 20\% distance uncertainty.  The relevant
$\chi^2$ values are 13.5
(best fit), 14.5 (COBE), 16.4 (no dipole), and 21.7
(LP).  The standard candle
frame would be consistent with COBE or zero motion, and inconsistent with LP,
but naturally with a lower certainty when using the poorer standard candle
assumption.  That is because there are real variations in the supernova
luminosities which are successfully accounted for by LCS.  Previous attempts to
use SN Ia as standard candles to measure motions of the Local Group have
probably suffered from poor photographic data and contained samples which were
too nearby to avoid local flows (Jergen \& Tammann 1993, Miller \& Branch
1992).   
  
\section{Errors, sample bias, and robustness}

The method of
$\Delta\chi^2$ intervals also predicts uncertainties on the cartesian velocity
components which result in ${\bf V} = (-90\pm370,-510\pm510,710\pm220$) km
s$^{-1}$.  Here the errors are 68\% confidence limits on each Cartesian
component considered separately (i.e. when $\chi^2$ is minimized with respect
to the other components).  Because supernovae are discovered at high galactic
latitude, our measurement of the motion is better in the z-direction which
points out of the galactic plane than in the x and y directions along the
galactic plane.  The LP frame predicts motion in the z-direction with the
opposite sign at $V_z$=$-333$ km s$^{-1}$, and this difference accounts for our
strong statistical rejection of that motion.  If we were to treat our single
estimate of the dipole as coming from the distribution defined by our errors,
then evaluating the best direction and magnitude of our vector would cause a
bias on the best-fit spherical coordinates.  For example, we note that the
maximum likelihood scalar length of our dipole, 600 km s$^{-1}$, is smaller
than the 880 km s$^{-1}$ which one would naively derive from the Cartesian
components (biased to higher values by the uncertainties, see Kendall 1976).

  To evaluate the reliability of our uncertainty estimates, we
performed a Monte Carlo simulation with 100,000 synthetic data sets of 13
supernovae with the same positions and uncertainties as ours, as displayed in
Figure 2.  These simulations verify our
$\Delta\chi^2$  uncertainty estimates
to within 1\% and demonstrate our ability to discriminate systematic motions on
the basis of 13 accurate distance indicators of wide spatial distribution.  
Our most likely agreement with COBE is at the 1 sigma level; the most likely
rejection of the LP frame is at the 3 sigma level.    In recovering our dipole,
we find a small geometric bias of (9,$-$28,$-$3) km s$^{-1}$ incurred by
simultaneously solving for the expansion and the motion of a sample which is
not uniformly distributed over the sky.       
 
We also performed a series of
jackknife tests to assess the influence of outliers on our results, with the
conclusion that our results are robust against deleting any 1 or 2 supernovae
from the sample. For example, removing the two farthest supernovae (new
effective velocity = 5,900 km sec$^{-1}$), the two nearest supernovae (new
effective velocity = 8,200 km sec$^{-1}$) or the two supernovae with largest
distance uncertainty had no significant effect on any of the results.   

Our
corrections for galactic absorption (Burstein and Heiles 1984) are small, and
omitting these corrections does not alter our result.  Host absorption appears
insignificant in our sample.  All of the supernovae were discovered far from
the centers of their galaxies and none of the spectra of these supernovae
showed strong NaI D absorption.  Finally, any error incurred by assuming the
sample mean host extinction is equal to that of SN 1972E affects only our
measurement of H$_o$, but not the direction of the inferred dipole.  Including
estimates for individual object deviation from the mean host extinction (up to
$\sigma_{A_V}$=0.15 mag) does not significantly alter our results.

Finally, we considered the effects of sparse sampling and clustering on our
ability to resolve a dipole.  Measurement of a dipole in a complex velocity
field from a limited number of objects can be distorted by small-scale
velocity patterns (Kaiser 1988, Feldman \& Watkins, 1994).   Uro\v{s} Seljak
(private communication) has calculated the window function for our sample and
convolved it with a standard CDM power spectrum to determine the components of
the covariance matrix and thus estimate the dipole components and their
uncertainty.  This procedure yields a maximum likelihood estimate of
$(-42\pm445,-517\pm575,716\pm276)$ km s$^{-1}$ which is highly consistent with
the value and uncertainty we derive for the dipole without considering these
effects.  Including velocity correlations also yields a greater consistency
with the COBE dipole ($\Delta\chi^2=2.4$) and a reduced lower bound on the
rejection of the LP frame (99.8\%) and the LP measurement (98.4\%).  The latter
is a lower bound because the two samples measure
 some of the same space (e.g.
SN 1993ae is in Abell 194) and have some
 small-scale flows in common.  We
intend to pursue N-body simulations to see
 how SNIa can be used to
discriminate among cosmological models (Seljak et al.
 1995).

\section{Interpretation}

Our result is consistent with the COBE rest frame,
but not with that proposed by LP.  Although our effective velocity is 7,000 km
s$^{-1}$ and the Abell Cluster inertial frame used by LP is effectively in the
range of 8,000-11,000 km s$^{-1}$, we can easily extend our result up to 8,000
km s$^{-1}$ by excluding two nearby supernovae.  In addition, LP have reduced
the effective velocity of their sample to 5,000 km s$^{-1}$ by excluding
distant clusters.  Neither change affects the two derived motions, or brings
them into agreement. 

The motion of the Local Group relative to the galaxies
in our sample is, within the errors, the same as our motion relative to the
CMB.  The most economical conclusion to draw is that the galaxies at 7,000 km
s$^{-1}$ are at rest in the cosmic frame.  If this is true, it is possible that
the LP measurement may be telling us something about the uniformity of the
galactic luminosity function on large scales.  With an enlarged supernova
sample, we will attempt to confirm and further constrain these interesting
results.  

\bigskip          

David Spergel audaciously suggested that our
small sample might constrain the large scale flow.  Brian Schmidt and George
Rybicki contributed valuable discussions.  We are grateful to Mario Hamuy, Mark
Phillips, Nick Suntzeff, Bob Schommer, Jos\`{e} Maza and the entire
Cal\'{a}n/Tololo collaboration for the opportunity to study their outstanding
data. Also our thanks to Uro\v{s} Seljak for providing large scale structure
calculations.  Cerro Tololo Inter-American Observatory, National Optical
Astronomy Observatories, is operated by the Association of Universities for
Research in Astronomy, (AURA), under cooperative agreement with the National
Science Foundation.  At Harvard, this work was suppported through grants
AST-92-18475 and PHY-91-06678.
 
\begin{table}[bp]     
\begin{center}              
\caption{SN Ia Parameters} \vspace{0.4cm}
\begin{tabular}{cccccc} \hline
   {\em Supernova}     &   $ \log v (km s^{-1})$       &  $l$ & $b$ & $\mu$ (LCS)$^*$ & $\sigma_{\mu}$    \\ 
 \hline \hline     
  1992bo & 3.753 & 261.88\deg & $-80.35$\deg &  34.53 & .08  \\
  1992bc & 3.782 & 245.70\deg & $-59.64$\deg &  34.48 & .08  \\
  1992K &  3.451 & 306.28\deg & 16.31\deg &  33.15 & .38  \\
  1992aq & 4.485 & 1.78\deg & $-65.32$\deg & 38.21 & .08  \\
  1992ae & 4.350 & 332.7\deg & $-41.99$\deg & 37.68 & .09  \\
  1992P & 3.872 & 295.62\deg  & 73.11\deg & 35.39 & .05  \\
  1992J & 4.120 & 263.55\deg  & 23.54\deg  &36.88 & .13   \\
  1991U & 3.968 & 311.82\deg  & 36.21\deg & 36.02 & .13   \\
  1991ag & 3.618 & 342.56\deg  & $-31.64$\deg &33.90 & .06  \\
  1990af & 4.178 & 330.82\deg  & $-42.24$\deg & 36.62 &.06  \\
  1992G & 3.257 & 184.62\deg & 59.84\deg & 32.57 & .08  \\
  1991M & 3.396 & 30.39\deg & 45.90\deg & 33.11 & .17 \\
  1993ae & 3.769 & 144.62\deg & $-63.23$\deg & 34.54 & .14  \\
\hline \hline
\multicolumn{6}{l}{*$\mu$ is our distance modulus; $\sigma_{\mu}$ is the uncertainty; log v from CTIO}
\end{tabular}
\end{center}
\end{table}
  
\def\refitem{\par\parskip 0pt\noindent\hangindent 20pt}
\newpage
\centerline {\bf References}
\vskip 12 pt    
\refitem  Avni, Y. 1976, ApJ, 210, 642
\\
\refitem  Burstein, D.  1990, Rep. Prog. Phys. 53,421
\\
\refitem Burstein, D., \& Heiles, C. 1982, AJ, 87, 1165
\\
\refitem  de Vaucouleurs, G. et al 1991, in {\it  Third Reference Catalogue of Bright Galaxies} (Springer-Verlag, New York)
\\
\refitem  Feldman, H.A. \& Watkins, R. 1994, ApJ, 430,L17 
\\
\refitem Ford, C. et al 1993, AJ, 106, 3  
\\
\refitem Hamuy, M., Phillips, M.M., Maza, J., Suntzeff, N.B., Schommer, R.A., \& Aviles, A.  1995b,in preparation
\\
\refitem Hamuy, M., Phillips, M.M., Maza, J., Suntzeff, N.B., Schommer, R.A., \& Aviles, A. 1995a, AJ, 000, 000
\\
\refitem Hamuy, M., Phillips, M.M., Maza, J., Suntzeff, N.B., Schommer, R.A., \& Aviles, A. 1994, AJ, 000, 000
\\  
\refitem Hamuy, M., et al 1993a, AJ, 106, 2392
\\
\refitem Hamuy, M., Phillips, M., Wells, L., \& Maza, J. 1993b, PASP, 105, 787
\\
\refitem  Jerjen, H. \& Tammann, G.A. 1993, A\&A, 276, 1
\\
\refitem Kaiser, N. 1988, MNRAS, 231, 149
\\
\refitem Kendall, M, \& Stuart, A., 1976, {\it The Advanced Theory of Statistics, Vol. III}, p. 97
\\
\refitem  Lauer, T.R. \& Postman, M. 1991,ApJ, 425,418
\\
\refitem  Lynden-Bell, D. \& Lahav, O. 1988, {\it Large-Scale Motions in the Universe} (ed. Rubin, V.C. \& Coyle, G.V.) (Princeton, Princeton University Press)
\\
\refitem  Marzke, R.O., Geller, M.J., daCosta, L.N. \& Huchra, J.P. 1995, AJ, submitted 
\\
\refitem Maza, J., Hamuy, M., Phillips, M., Suntzeff, N., Aviles, R. 1994, ApJ, 424, L107
\\
\refitem  Miller, D.L., Branch, D. 1991,AJ, 103, 379
\\
\refitem Phillips, M. 1993, ApJ, 413, L105
\\
\refitem  Press, W.H., Teukolsky, S.A., Vetterling, W.T. \& Flannery, B.P. 1992, {\it Numerical Recipes, 2nd ed.} (Cambridge University Press)
\\
\refitem  Riess, A.G., Press W.H., Kirshner, R.P. 1995, ApJ, 000, L00
\\  
\refitem Riess, et al, 1995 (in preparation)   
\\
\refitem Rowan-Robinson, M. et al, 1990, MNRAS, 247, 1
\\
\refitem  Rubin, V.C., Thonnard, N., Ford, W.C. \& Roberts, M.S. 1976, AJ, 81, 687
\\
\refitem  Rubin, V.C., 1977, ApJ, 211,L1 
\\
\refitem  Sandage, A. et al., 1994, ApJ, 423, L13
\\
\refitem Seljak, Uro\v{s} et al. 1995, in preparation
\\
\refitem  Smoot, G.F. et al. 1992, 396, L1 
\\
\refitem  Strauss, M.A., Ostriker, J.P., Cen, R., Lauer, T.R. \& Postman, M., 1995, ApJ, 000, 000  

\newpage

Figure 1: Hubble diagram velocity residuals and predicted Local Group
motion on the sky.  Filled/Open points represent supernovae with
negative/positive residual velocity; the area of the points correspond
to the magnitude of the velocity residual.  Filled/Open crosses show
the direction towards which the Local Group is approaching/receding
according to the best fit for the data in this paper, the COBE
satellite, and Lauer and Postman's survey of brightest cluster
galaxies.  One and two sigma contours are displayed for the direction
of the best fit to the supernova residual velocities. Coordinates are
galactic with the central meridian at 335$\deg$.

\bigskip

Figure 2: Monte Carlo dipole recovery simulation.  100,000 synthetic
data sets of 13 supernovae were created with the same positions and
uncertainties as our sample and our adopted dipole motion of
$(-90,-510,710)$ km s$^{-1}$.  These are the distributions of the
cartesian components of the dipole recovered from each synthetic data
set.  The difference between the mean (solid line) and the adopted
dipole (dotted line) represents a geometric bias of $(9,-28,-3)$ km
s$^{-1}$.  The 1 $\sigma$ boundaries (outer solid lines) confirm the
$\chi^2$ errors to within 1\%.  The arrows show the COBE and LP
measurement of the dipole components.  This simulation also verified
our most likely agreement with COBE to be at the 1 $\sigma$ level, the
most likely disagreement with LP to be at the 3 $\sigma$ level.  The
difference between our result and the LP result is most conspicuous in
the z-direction.

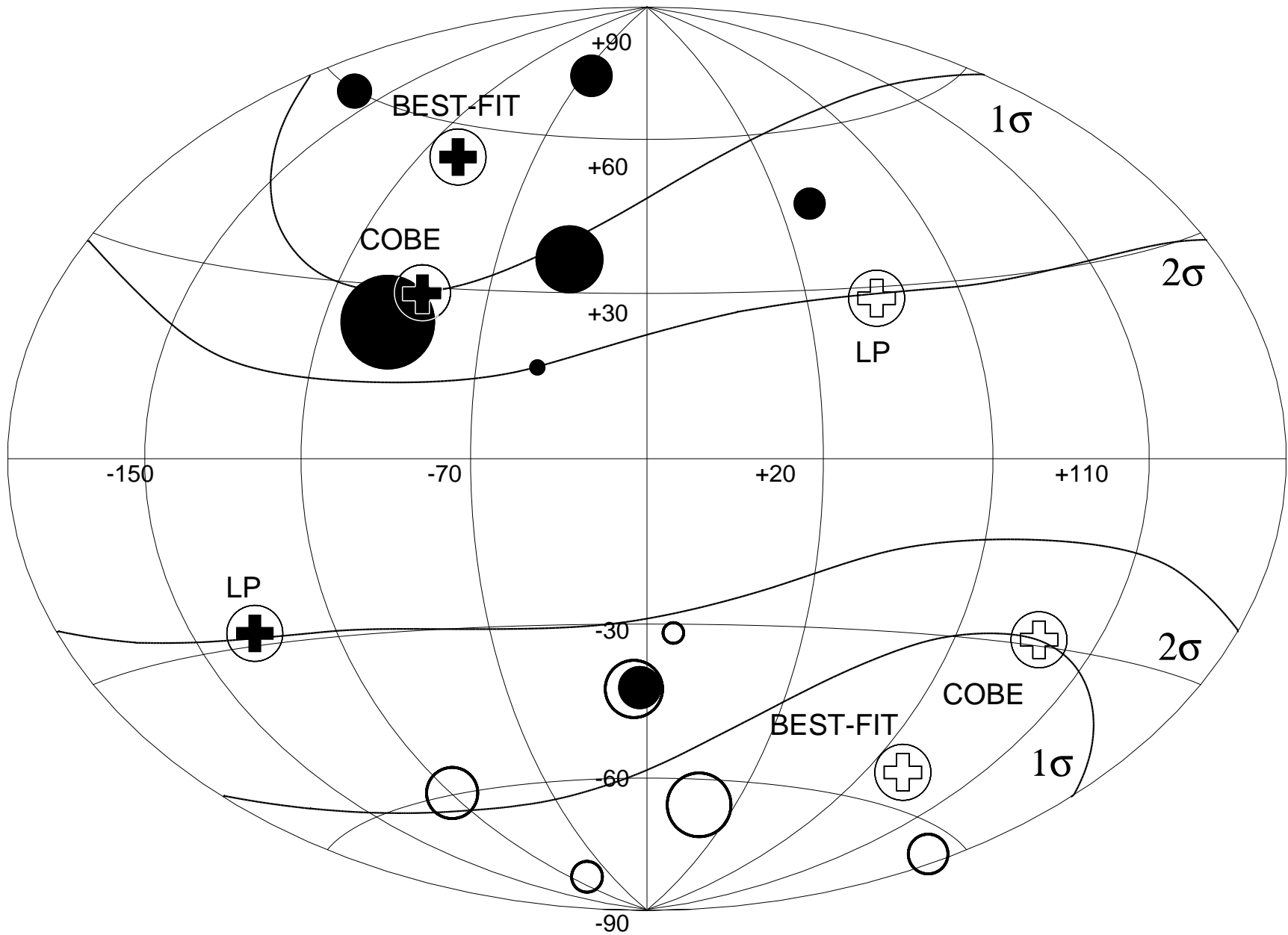
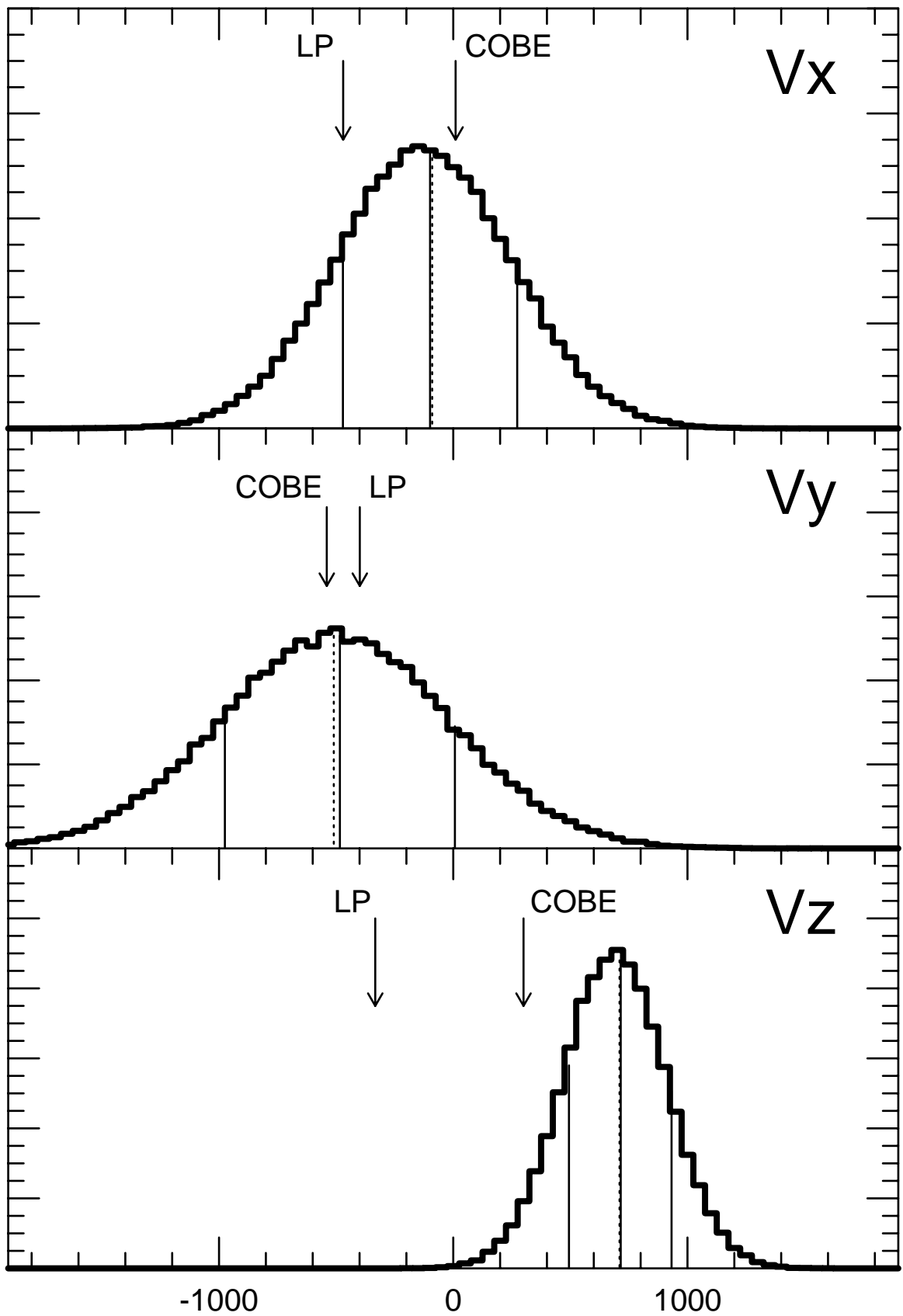
\end{document}